\documentclass[a4paper,preprint,showkeys]{amsproc}
\usepackage{mathrsfs}
\usepackage[utf8]{inputenc}
\usepackage{mathtools}
\usepackage{amsmath,booktabs}
\usepackage[T4, OT1]{fontenc}
\usepackage[arrow, matrix, curve]{xy}
\usepackage{newunicodechar}
\usepackage{graphicx}
\usepackage{bm}
\usepackage[dvips]{epsfig}
\usepackage{rotating}
\usepackage{amsmath,amssymb,amsfonts,amsthm}
\usepackage{amsmath,amscd}
\usepackage{subfig}
\usepackage{caption}
\usepackage{pst-all}
\usepackage{dcolumn}
\usepackage[thinc]{esdiff}
\usepackage{amsthm}
\usepackage{dsfont}
\usepackage{amssymb}
\usepackage[hyphens]{url} 
\usepackage[dvips]{graphicx}
\usepackage{xparse}
\usepackage{physics}
\usepackage{mathrsfs}
\makeatletter
\@namedef{subjclassname@2024}{\textup{2024} Mathematics Subject Classification}
\makeatother
\theoremstyle{plain}

\theoremstyle{definition}

\theoremstyle{remark}
 
 \numberwithin{equation}{section}

\renewcommand{\leq}{\leqslant}\renewcommand{\geq}{\geqslant}

\newcommand{\ie}{i.\,e.\ }

\title[Exact shape of an ideal heavy spring suspended at both ends.]{Exact shape of an ideal heavy spring\\ suspended at both ends.}
\author[Belardinelli]{Cyril Belardinelli\\Kantonsschule Solothurn\\Switzerland} 
\address{ 
Kantonsschule Solothurn\\
Solothurn\\
Switzerland}
\email{cyril.belardinelli@ksso.ch}
\begin{document}
\vspace{18mm} \setcounter{page}{1} \thispagestyle{empty}
\begin{abstract}
In the present article we determine the exact shape of an ideal spring suspended at both ends. Historically, this goes under the name of \textit{Elastica} problem.  The total energy of the suspended spring is a functional depending on two independent functions: on one hand the function describing the shape of the hanging spring, on the other hand a non-negative function describing the mass distribution along the elongated spring. Calculus of Variations provides a rather short and elegant solution. This method does not appear to be widely known in the literature.
\end{abstract}
\maketitle
\section{\label{history}Historical Context}
\begin{figure}[h!]
\centering
{\includegraphics[width=0.5\linewidth, angle=0]{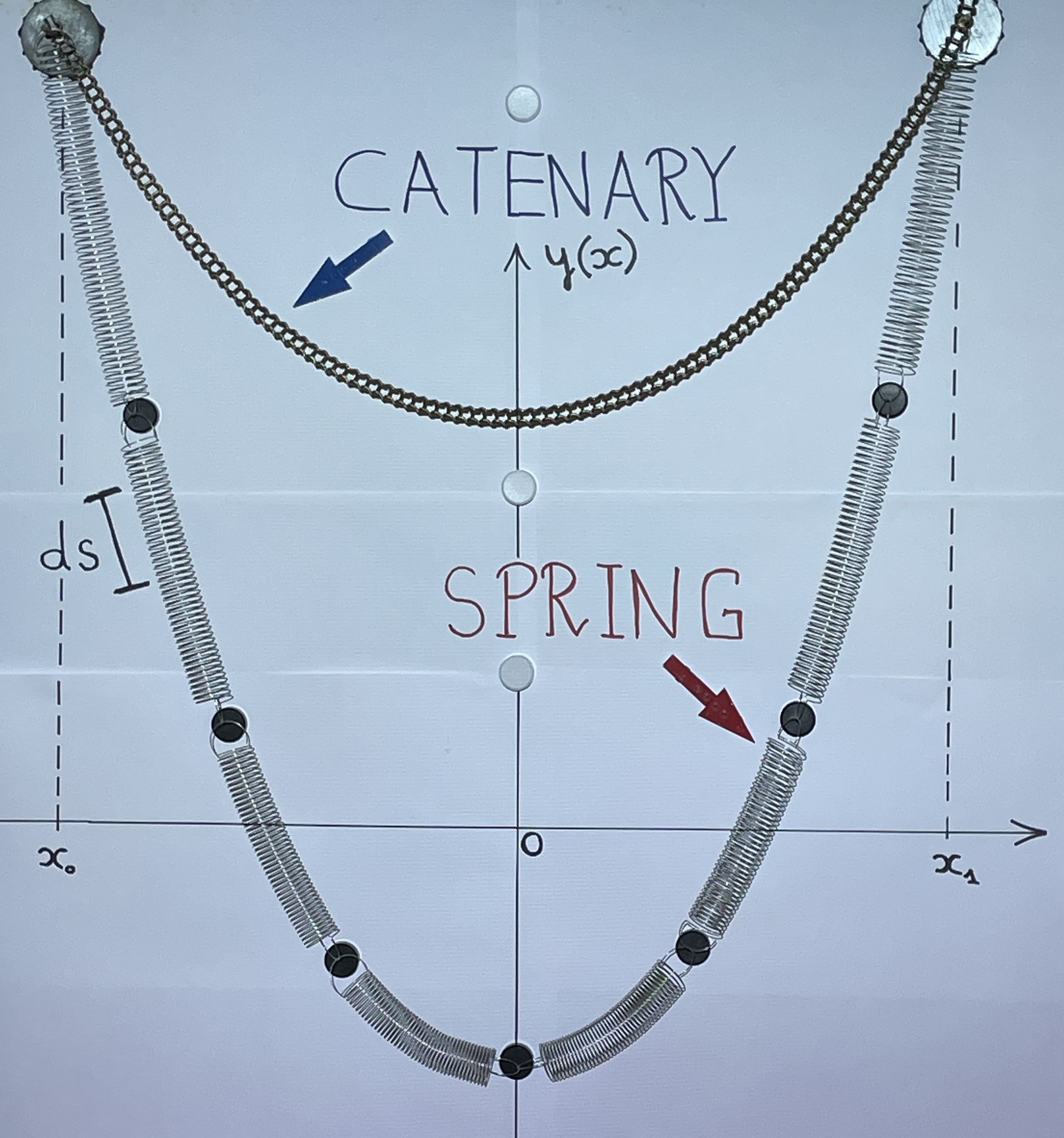}}
\caption{hanging unextendable catenary (top) and elastic spring (bottom) suspended at both ends. The spring's equilibrium length $\ell_{0}$ is identical to the catenary's invariable length. The entire spring is composed of identical shorter springs. (between black markings)\\Source: 
Photo taken at \textit{Kantonsschule Solothurn (Physics Lab)}}
\label{fig1}
\end{figure}
Catenary, the curve of a hanging chain, has been studied by many Renaissance and early modern mathematicians including Huygens, Leibniz and Bernoulli brothers. The catenary could be tackled with the newly invented calculus at the end of $17\text{th}$ century. Before the advent of the new calculus many believed the curve is a parabola\cite{kline:1991, boyer:1968}. In 1638, Galileo Galilei (1564-1642) treated the catenary in his famous \textit{Discorsi}\footnote{Numerous historians of Science attribute to Galileo the erroneous claim that the catenary curve is a parabola\cite{kline:1991, boyer:1968}. This is too one-sided a claim. True, in the second day of \textit{Discorsi} he asserts, through the voice of Salviati, that the catenary assumes the shape of a parabola. However, in the fourth day, he explains in plain words that the parabola is only an approximation of the catenary.\cite{galilei:1638}}\cite{galilei:1638}.   
Jakob Bernoulli (1654-1705) was among the first to use the new calculus to analytically solve differential equation problems.
In 1690, he posed the problem of finding the shape of a hanging elastic rope. Jakob does not specify any condition which limits the problem to the nonelastic case\cite{alassi:2020}. The so-called \textit{Elastica} problem he raises is therefore more general than the \textit{Catenaria} problem. 
\\
He invited leading European mathematicians to take part in
the challenge to which Gottfried Leibniz immediately responded.
In the \textit{Acta Eruditorium} of June 1691, Leibniz, Huygens, and Johann Bernoulli independently published their solutions.\\
Johann Bernoulli was immensely proud of having been able to solve the catenary problem\footnote{In a letter to Pierre Rémond de Montmort (1678-1719), dated 29 septembre 1718 he writes: \textit{... je Vous etonne, dites Vous, de dire que mon Frere n’a pu resoudre le probleme de la chainette; oui, je Vous le dis encore, car c’est une verité incontestable...,; Vous dites que mon frere avoit proposé ce probleme; cela est vrai, mais s’en suit il que dans ce tems il en avoit une solution? point du tout: lorsqu’il proposa ce probleme par ma suggestion (car j’y ai pensé le premier), nous n’etions encore ni l’un ni l’autre en etat de le resoudre, nous desesperions meme comme d’une chose insoluble, jusqu’à ce que Mr. Leibnits eut averti le public dans les Actes de Leipsic de 1690, p. 360, d’avoir resolu ce probleme sans donner la solution pour laisser du tems aux autres Analystes, ce qui nous anima mon frere et moi à nous y appliquer de nouveau; les efforts de mon frere furent sans succes, pour moi je fus plus heureux car je trouvai l’adresse (je le dis sans me vanter, pourquoi cacherois-je la verité?) de le resoudre pleinement et de le reduire à la rectification de la parabole, il est vrai que cela me couta des meditations qui me deroberent le repos d’une nuit entiere, c’etoit beaucoup pour ce tems là et pour le peu d’age et d’exercice que j’avois, mais le lendemain tout rempli de joie je courus ches mon frere, qui luttoit encore miserablement avec ce noeud Gordien sans rien avancer, soupçonnant toujours comme Galilée que la chainette etoit une parabole; Cessés! cessés! lui dis-je, ne Vous tourmentés plus à chercher l’identité entre la chainette et la parabole là où il n’y en a point... }}, while his brother, who had proposed it, had apparently failed to do so\cite{kline:1991}.
However, this judgment must be taken with a grain of salt; there was fierce competition between the two brothers. Jacob was mainly interested to provide a general solution for the \textit{Elastica} problem. In fact, recent research has shown that Jakob Bernoulli solved the problem of a hanging elastic rope in a more general context by unifying the \textit{catenary} with the \textit{elastica} problem\footnote{Quote from \cite{alassi:2020}:\textit{...J. Bernoulli had successfully unified the catenary problem and
the elastica problem. This goal was rooted in the challenge he started in 1690. But he did not publish his results, and his success was not perceived at the time, nor by modern historians.} }. A very detailled account of these new findings can be found in \cite{alassidiss:2020} and \cite{alassi:2020}. However, Jakob did not publish his results. We know about his achievements because he recorded them in his private notebook \textit{Meditationes}.\footnote{\textit{Meditationes, Annotationes, Animadversiones Theologicae \& Philosophicae, a me JB.concinnatae \& collectae ab anno 1677} –conserved at the Manuscript Department of Basel University Library.{\cite{alassi:2020}}}. The problem of unifying the catenary with the elastica problem has been revisited almost two decades after Jakob's death by Leonhard Euler (1707-1783) and Daniel Bernoulli (1700-1782) apparently completely unaware of Jakob's achievements\cite{alassidiss:2020}. Elastic chains were also a topic of interest in the 19th century\cite{finck:1826}.
\section{\label{intro}The \textit{Elastica} Problem revisited by Calculus of Variations}
The catenary is the solution to the problem of minimizing the potential gravitational energy of an unextendable (but flexible) chain hanging under its own weight when supported only at
its ends. In the present paper we study a natural extension of the problem which historically goes under the name of \textit{Elastica}. Instead of a flexible but unextendable chain we consider an elastic spring hanging under its own weight (See Fig.~\ref{fig1}).\\ In this case the total energy stored in the spring is given by the sum of the potential gravitational energy $E_{\text{gravity}}$ and the elastic energy $E_{\text{elastic}}$ due to the deformation of the spring. The shape of the hanging spring (in its equilibrium position) will be the one that minimizes the total energy.  It is therefore very natural to study the problem in the context of Calculus of Variations.\\ 
The total energy is a functional $E_{\text{total}}[y, \sigma]$ depending on two independent functions $y(x)$ and $\sigma(x)$ where the function $y(x)$ describes the shape of the hanging spring and $\sigma(x)$ denotes the mass distribution along the hanging spring defined as $\sigma(x)=d{m}/d{s}$ where $x$ is a point on the abscissa  and $d{s}$ is an infinitesimal line element of the spring (Fig.~\ref{fig1}).\\
The spring's mass is $m$ and its elastic constant is denoted by $k$. The unstretched spring has an equilibrium length $\ell_{0}$ and a mass density $\sigma_{0}=m/\ell_{0}$.\\The total energy functional of the spring is then formally;
\begin{eqnarray}
\label{functional}
E_{\text{total}}[y, \sigma]=E_{\text{gravity}}[y, \sigma]+E_{\text{elastic}}[y, \sigma]
\end{eqnarray}
The gravitational potential energy is given by the following integral;
\begin{eqnarray*}
E_{\text{gravity}}[y, \sigma]&=&g\int_{x_{0}}^{x_{1}} y(x)\, \sigma(x) \sqrt{1+y^{\prime 2}}\, d{x}
\end{eqnarray*}
Whereas the functional of the elastic energy can be formulated in the following way. (For a derivation see App.~\ref{app_A})
\begin{eqnarray*}
E_{\text{elastic}}[y, \sigma]&=&\frac{k \ell_{0}}{2}\int_{x_{0}}^{x_{1}}\qty[\frac{\sigma_{0}}{\sigma(x)}-2+\frac{\sigma(x)}{\sigma_{0}}] \sqrt{1+y^{\prime 2}}\, d{x}
\end{eqnarray*}
We must minimize the energy functional of Eq.~\ref{functional} under the constraint of fixed mass $m$. In addition, we demand the linear mass density $\sigma(x)$ to be non-negative.
\begin{eqnarray*}
m=\int_{x_{0}}^{x_{1}}\sigma(x) \sqrt{1+y^{\prime 2}}\, d{x}\quad \text{and} \quad \sigma(x) \geq 0\quad \forall x \in [x_{0},x_{1}]
\end{eqnarray*}
In order to take into account the first of these constraints one introduces a \textit{Lagrange-Multiplier} $\mu$ which leads to the functional $H[y,\sigma]$;
\begin{eqnarray*}
H[y,\sigma]=E_{\text{total}}[y, \sigma]+\mu \int_{x_{0}}^{x_{1}}\qty[\sigma(x) \sqrt{1+y^{\prime 2}}\, -m]d{x}\equiv \int_{x_{0}}^{x_{1}}\mathcal{F}(x)\, d{x}
\end{eqnarray*}
where the equivalence "$\equiv$" means: "equal up to irrelevant constant terms".\\
The Lagrange density $\mathcal{F}(x)$ is;
\begin{eqnarray*}
\mathcal{F}(x)=\qty[g y(x)\sigma(x)+\mu \sigma(x) +\frac{k \ell_{0}\sigma_{0}}{2\sigma(x)}+\frac{k \ell_{0}\sigma(x)}{2\sigma_{0}}-k \ell_{0}]\sqrt{1+y^{\prime 2}}
\end{eqnarray*}
The variation of H with respect to $y$  and $\sigma$ leads to two coupled differential equations; \ie the \textit{Euler-Lagrange equations};
\begin{eqnarray*}
\dv{x} \pdv{F}{y^{\prime}}-\pdv{F}{y}=0\\
\dv{x} \pdv{F}{\sigma^{\prime}}-\pdv{F}{\sigma}=0
\end{eqnarray*}
An explicit evaluation leads to the following system of equations;
\begin{eqnarray*}
\dv{x}\qty{\qty[g\sigma (y+\mu/g)+\frac{k \ell_{0}}{2}\qty(\frac{\sigma_{0}}{\sigma}+\frac{\sigma}{\sigma_{0}}-2)]\frac{y^{\prime}}{\sqrt{1+y^{\prime 2}}}}&=&g\sigma \sqrt{1+y^{\prime 2}}\\
\alpha y+\frac{2\mu \sigma_{0}}{k \ell_{0} }+\qty(1-\frac{\sigma_{0}^2}{\sigma^{2}})&=&0
\end{eqnarray*}
where we introduced the parameter $\alpha$ defined as;
\begin{equation*}
\alpha=\frac{2 g\sigma_{0}}{k \ell_{0}}
\end{equation*}
Note that the parameter $\alpha$ has dimension $[\alpha]=\text{m}^{-1}$.\\
We rewrite the system by introducing the following variables;
\begin{eqnarray*}
z&:=&y+\mu/g\\
\tilde{\sigma}&:=&\frac{\sigma}{\sigma_{0}}
\end{eqnarray*}
which gives;
\begin{eqnarray}
\dv{x}\qty{\qty[(1+\alpha z)\tilde{\sigma}+\frac{1}{\tilde{\sigma}}-2]\frac{z^{\prime}}{\sqrt{1+z^{\prime 2}}}}&=&\alpha\tilde{\sigma} \sqrt{1+z^{\prime 2}}\\
1+\alpha z&=&\frac{1}{\tilde{\sigma}^2}
\end{eqnarray}
Elimination of the reduced density $\tilde{\sigma}$ from the first equation (I) leads to a $2\text{nd}$ order differential equation for the variable $z$ alone;
\begin{equation}
\label{dimensional_de}
z^{\prime \prime}\qty(1+\alpha z-\sqrt{1+\alpha z})=\frac{\alpha}{2}\qty(1+z^{\prime 2})
\end{equation}
It is convenient to cast the differential equation in a dimensionless form. For that purpose we choose the variables;
\begin{eqnarray}
\tilde{z}&=&1+\alpha z\\
\tilde{x}&=&\alpha x
\end{eqnarray}
We get then the following \textit{universal} differential equation (DE);
\begin{equation}
\label{de}
\ddot{\tilde{z}}\qty(\tilde{z}-\sqrt{\tilde{z}})=\frac{1+\dot{\tilde{z}}^{2}}{2}
\end{equation}
We can draw an important conclusion from the latter differential equation: On one hand it is evident from physical reasons that $\ddot{\tilde{z}}(0)>0$. On the other hand we see that the r.h.s. of Eq.~\ref{de} is $\geq \frac{1}{2}$. We therefore conclude that $\tilde{z}(0)>1$.\\
Since $\tilde{z}=\frac{1}{\tilde{\sigma}}$ applies and the linear density $\tilde{\sigma}$ assumes its lowest value at $\tilde{x}=0$, we have necessarily;
\begin{equation*}
\tilde{z}(\tilde{x})>1\iff \tilde{\sigma}(\tilde{x})<1 \quad \forall \, \tilde{x}\in \mathbb{R}
\end{equation*}
Moreover, we conclude from Eq.~\ref{de};
\begin{equation*}
\lim _{\tilde{z}(0) \to 1^{+}}\ddot{\tilde{z}}(0)=+\infty
\end{equation*}
The limit $\ddot{\tilde{z}}(0) \rightarrow +\infty $ means that the radius of curvature  in $\tilde{x}=0$ (\ie in the lowest point of the curve $\tilde{z}(\tilde{x})$ where $\dot{\tilde{z}}(0)=0$) is approaching zero (\ie infinite curvature). This happens when the distance between the  two suspension points of the spring goes to zero. (In the limit where they coincide one would have two vertically and infinitesimally adjacent hanging springs connecting at the bottom) 
\subsection{Solving the differential equation}
In the present section we provide an exact solution of the following Initial value problem (IVP) ;
\begin{equation}
\label{dimless}
\begin{split}
\ddot{\tilde{z}}\qty(\tilde{z}-\sqrt{\tilde{z}})&=\frac{1+\dot{\tilde{z}}^{2}}{2}\\
\tilde{z}(0)&=\tilde{z}_{0}\\
\dot{\tilde{z}}(0)&=0
\end{split}
\end{equation}
First, we introduce a new variable $p$;
\begin{equation*}
p(z)= \dv{\tilde{z}}{\tilde{x}}
\end{equation*}
With this variable we can write;
\begin{equation*}
\ddot{\tilde{z}}=\dot{p}p
\end{equation*}
The differential equation Eq.~\ref{dimless} transforms into;
\begin{equation*}
\log{\qty(1+p^2)}+C=\int_{\tilde{z}_{0}}^{\tilde{z}} \frac{d{\tilde{z}^{\prime}}}{\tilde{z}^{\prime}-\sqrt{\tilde{z}^{\prime}}}
\end{equation*}
The latter integral can be calculated;
\begin{equation*}
\int_{\tilde{z}_{0}}^{\tilde{z}} \frac{d{\tilde{z}^{\prime}}}{\tilde{z}^{\prime}-\sqrt{\tilde{z}^{\prime}}}=2\log{\underbrace{\frac{r-1}{r_{0}-1}}_{>1}}>0
\end{equation*}
where
\begin{eqnarray*}
r&=&\sqrt{\tilde{z}}\\
r_{0}&=&\sqrt{\tilde{z}_{0}}
\end{eqnarray*}
By applying the initial value $p(0)=\dot{\tilde{z}}(0)=0$ we conclude that $C=0$. \\
\\With this we obtain;
\\
\begin{equation*}
p(z)= \dv{\tilde{z}}{\tilde{x}}=\pm\sqrt{\qty(\frac{r-1}{r_{0}})^2-1}
\end{equation*}\\
This leads to a further integration;
\begin{equation*}
\pm\int_{r_{0}}^{r}{\frac{2r d{r}}{\sqrt{\qty(\frac{r-1}{r_{0}})^2-1}}}=\int_{0}^{\tilde{x}}{d{\tilde{x}^{\prime}}}
\end{equation*}
By the variable change $\eta=\frac{r-1}{r_{0}}$ the latter equation transforms to;
\begin{equation*}
2\qty(r_{0}-1)\int_{1}^{\eta}{\frac{d{\eta^{\prime}}}{\sqrt{\eta^{\prime 2}-1}}}+\qty(r_{0}^2-1)\int_{1}^{\eta}{\frac{2\eta^{\prime}d{\eta^{\prime}}}{\sqrt{\eta^{\prime 2}-1}}}=\pm\tilde{x}
\end{equation*}
These integrals are readily evaluated. the result is;
\begin{equation}
\label{exact_solution}
\pm\tilde{x}=2\qty(r_{0}-1)^{2}\sqrt{\eta^2-1}+2\qty(r_{0}-1)\cosh^{-1}{\eta}
\end{equation}
where 
\begin{equation*}
\eta=\frac{r-1}{r_{0}-1}=\frac{\sqrt{\tilde{z}}-1}{\sqrt{\tilde{z}_{0}}-1}
\end{equation*}
The inverse of hyperbolic cosine can be expressed in the following way;
\begin{equation*}
\cosh^{-1}(\eta)=\log(\eta+\sqrt{\eta^2-1})\qquad 1\leq \eta < \infty
\end{equation*}
Therefore, we may also write;
\begin{equation*}
\pm\tilde{x}=2\qty(r_{0}-1)^{2}\sqrt{\eta^2-1}+2\qty(r_{0}-1)\log(\eta+\sqrt{\eta^2-1})
\end{equation*}
\begin{figure}[h!]
  \centering
  \subfloat[][]{\includegraphics[width=0.4\linewidth, angle=90]{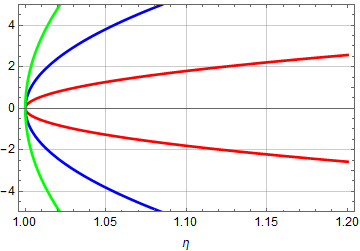}}
\qquad
\subfloat[][]{\includegraphics[width=0.4\linewidth, angle=90]
{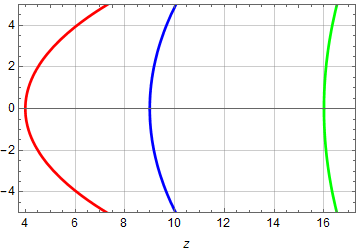}}
\caption{Graph $\eta=\eta(\tilde{x})$ (A) and graph of $\eta=\eta(\tilde{x})$ (B) describing the shape of elastic catenaries for $r_{0}=2.0$ (Red), $r_{0}=3.0$ (Blue), $r_{0}=4.0$ (Green)}
\label{fig2}
\end{figure}
\newpage
Note that the reduced density $\tilde{\sigma}$ is given by the formula;
\begin{equation*}
\tilde{\sigma}=\sqrt{\frac{1}{\tilde{z}}}
\end{equation*}
\begin{figure}[h!]
  \centering
{\includegraphics[width=0.6\linewidth]
{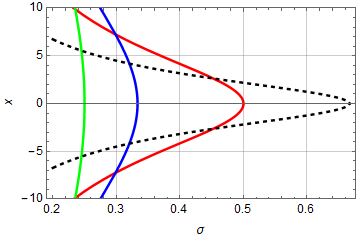}}
\caption{Graph $\pm\tilde{x}=\tilde{x}(\tilde{\sigma})$ describing the reduced mass density $\tilde{\sigma}$ for values $r_{0}=4.0$ (Green), $r_{0}=3.0$ (Blue), $r_{0}=2.0$ (red) and $r_{0}=1.5$ (dashed)}
\label{fig2}
\end{figure}
\subsection{Limiting cases of soft and hard springs}
in order to perform the extreme limits of hard and soft springs we have to go back to Eq.~\ref{dimensional_de}. For the sake of clarity we rewrite the equation here; 
\begin{equation*}
z^{\prime \prime}\qty(1+\alpha z-\sqrt{1+\alpha z})=\frac{\alpha}{2}\qty(1+z^{\prime 2})
\end{equation*}
\subsubsection{limiting case of a hard spring}
In this limit one must consider the limit;
\begin{equation*}
\alpha=\frac{2 g\sigma_{0}}{k \ell_{0}}\rightarrow 0.
\end{equation*}
By using the following approximation (Taylor expansion up to $1\text{st}$ order); 
\begin{equation*}
\sqrt{1+\alpha z}=1+\frac{1}{2}\alpha z + \mathcal{O}(\alpha^2)
\end{equation*}
and by retaining terms up to first order in $\alpha$ DE Eq.~\ref{dimensional_de} becomes (after cancelling out $\alpha$);
\begin{eqnarray}
\label{dimensional_de_2}
z^{\prime \prime}z=\qty(1+z^{\prime 2})
\end{eqnarray}
By taking the derivative on both sides of Eq.~\ref{dimensional_de_2} we obtain;
\begin{eqnarray*}
& z^{\prime \prime \prime}z+z^{\prime \prime}z^{\prime}=2z^{\prime \prime} z^{\prime}\\
\iff & z^{\prime \prime \prime}z=z^{\prime \prime}z^{\prime}\\
\iff & z^{2} \qty(\frac{z^{\prime \prime}}{z})^{\prime}=0
\end{eqnarray*}
This leads to the solution $z_{1}(x)=0$ which is the trivial solution of Eq.~\ref{dimensional_de_2}. The non-trivial solution denoted $z_{2}(x)$ fullfills;
\begin{eqnarray*}
&\qty(\frac{z_{2}^{\prime \prime}}{z_{2}})^{\prime}=0\\
\iff & z_{2}^{\prime \prime}=c\cdot z_{2}
\end{eqnarray*} 
The \textit{catenary}, \ie the \textit{hyperbolic cosine}  is the well-known physically relevant solution to the latter equation.
\subsubsection{limiting case of a soft spring}
In this case we have to consider the limit;
\begin{equation*}
\alpha=\frac{2 g\sigma_{0}}{k \ell_{0}}\rightarrow \infty.
\end{equation*}
In this limit we get for Eq.~\ref{dimensional_de};
\begin{eqnarray*}
z^{\prime \prime}z=\frac{1+z^{\prime 2}}{2}
\end{eqnarray*}
Once again, we take the derivative on both sides;
\begin{eqnarray*}
& z^{\prime \prime \prime}z+z^{\prime \prime}z^{\prime}=z^{\prime \prime} z^{\prime}
\iff & z^{\prime \prime \prime}z=0\\
\end{eqnarray*}
Once again, we have a trivial solution $z_1(x)=0$. The non-trivial solution is determined by the equation;
\begin{equation}
\label{non-trivial}
z_{2}^{\prime \prime \prime}(x)=0
\end{equation}
A general solution to Eq.~\ref{non-trivial} is a parabola. The wrong conjecture that the catenary is identical to a parabola reappears here but only in the limit of a very soft spring (\ie in the limit $k \rightarrow 0$).
\begin{appendix}
\section{Elastic Energy}
\label{app_A}
The elastic energy of a spring, also known as elastic potential energy, is the energy stored within a spring when it is stretched or compressed from its equilibrium position. We assume a general non-homogeneous deformation. The stored energy is then a functional $E_{\text{elastic}}$ depending on the linear mass distribution $\sigma(x)$ along the spring.\\
To begin with we consider an infinitesimal line element $d{s}$ of the spring which stores an (infinitesimal) amount $d{E}_{\text{elastic}}$ of elastic energy;
\begin{equation*}
d{E}_{\text{elastic}}=\frac{1}{2}{k}_{0}\cdot (d{s}-d{s}_{0})^{2}
\end{equation*}
where $d{s}_{0}$ is the equilibrium length of $d{s}$ and ${k}_{0}$ is the spring's elastic constant of the infinitesimal spring (of length $d{s}_{0}$).
The elastic constant of $d{s}_{0}$ is;
\begin{equation*}
k_{0}=k\cdot \frac{\ell_{0}}{d{s}_{0}}\\
\end{equation*}
Since 
\begin{eqnarray*}
\sigma(x)d{s}=d{m}=\sigma_{0}d{s}_{0}
\iff d{s}_{0}=\frac{\sigma(x)}{\sigma_{0}}d{s} 
\end{eqnarray*}
we have;
\begin{equation*}
d{E}_{\text{elastic}}=\frac{1}{2}k\ell_{0}\frac{\sigma_{0}}{\sigma(x)}\qty(1-\frac{\sigma(x)}{\sigma_{0}})^{2}d{s}
\end{equation*}
Finally, with the following relation between $d{x}$ and $d{s}$;
\begin{equation*}
d{s}=d{x}\sqrt{1+y^{\prime 2}}
\end{equation*}
we obtain the total elastic energy in integral form;
\begin{equation*}
E_{\text{elastic}}=\frac{1}{2}k\ell_{0}\int_{x_{0}}^{x_{1}}
\frac{\sigma_{0}}{\sigma(x)}\qty(1-\frac{\sigma(x)}{\sigma_{0}})^2 \sqrt{1+y^{\prime 2}}\, d{x}
\end{equation*}
\end{appendix}
\bibliography{elastic_cathenary}
\bibliographystyle{plain}
\end{document}